\documentclass[conference]{IEEEtran}
\pdfoutput=1
\usepackage[colorlinks]{hyperref}

\usepackage{amsmath}
\usepackage{amsthm}
\usepackage{amssymb}
\usepackage{bm}
\usepackage{xspace}
\usepackage{xcolor}
\usepackage{graphicx}
\usepackage{url}
\usepackage{framed}
\usepackage{float}
\usepackage{rotating}
\usepackage{verbatim}
\usepackage{listings}
\usepackage{lscape}

\usepackage{pgfplots}
\usepackage{pgf}
\usepackage{tikz}
\usetikzlibrary{arrows,shapes.misc,chains,scopes}
\usetikzlibrary{matrix}
\usetikzlibrary{calc}
\usetikzlibrary{fit}

\usepackage{pgfplotstable}
\usepackage{booktabs}
\usepackage{colortbl}

\newcommand{\executeiffilenewer}[3]{%
\ifnum\pdfstrcmp{\pdffilemoddate{#1}}%
{\pdffilemoddate{#2}}>0%
{\immediate\write18{#3}}\fi%
}

\graphicspath{{figures/}}

\theoremstyle{plain}

\newtheorem{theorem}{Theorem}

\newcounter{algocount}
\newcounter{examplecount}

\newcommand{\veca}{\boldsymbol{a}}
\newcommand{\vecb}{\boldsymbol{b}}
\newcommand{\vecB}{\boldsymbol{B}}

\newcommand{\vecL}{\boldsymbol{L}}

\newcommand{\vecS}{\boldsymbol{S}}

\newcommand{\capacity}{\ensuremath{\mathsf{C}}\xspace}

\newcommand{\setb}{\ensuremath{\mathcal{B}}\xspace}
\newcommand{\setc}{\ensuremath{\mathcal{C}}\xspace}

\newcommand{\sett}{\ensuremath{\mathcal{T}}\xspace}

\newcommand{\setx}{\ensuremath{\mathcal{X}}\xspace}

\newcommand{\bmm}{\begin{matrix}}
\newcommand{\emm}{\end{matrix}}
\newcommand{\bpm}{\begin{pmatrix}}
\newcommand{\epm}{\end{pmatrix}}

\newcommand{\bsbm}{\left[\begin{smallmatrix}}
\newcommand{\esbm}{\end{smallmatrix}\right]}
\newcommand{\bspm}{\left(\begin{smallmatrix}}
\newcommand{\espm}{\end{smallmatrix}\right)}

\newcommand{\bbm}{\begin{bmatrix}}
\newcommand{\ebm}{\end{bmatrix}}

\DeclareMathOperator*{\argmin}{argmin}

\DeclareMathOperator{\expop}{\mathbb{E}}
\DeclareMathOperator{\entop}{\mathbb{H}}
\DeclareMathOperator{\miop}{\mathbb{I}}
\DeclareMathOperator{\kl}{\mathbb{D}}

\DeclareMathOperator*{\st}{subject\;to}

\newcommand{\oleq}[1]{\overset{\text{(#1)}}{\leq}}

\usepackage{cite}
\usepackage{printlen}

\title{Probabilistic Signal Shaping\\for Bit-Metric Decoding}

\author{\IEEEauthorblockN{Georg B\"ocherer}
\IEEEauthorblockA{Institute for Communications Engineering\\Technische Universit\"at M\"unchen, Germany\\
Email: \texttt{georg.boecherer@tum.de}}
\thanks{This work was supported by the German Ministry of Education and Research in the framework of an Alexander von Humboldt Professorship.}
}

\newcommand{\snr}{\ensuremath{\mathsf{SNR}}\xspace}

\newcommand{\perr}{P_\text{iw}}
\newcommand{\bits}{\boldsymbol{B}}

\graphicspath{{figures/}}

\begin{document}

\maketitle

\begin{abstract}
A scheme is proposed that combines probabilistic signal shaping with bit-metric decoding. The transmitter generates symbols according to a distribution on the channel input alphabet. The symbols are labeled by bit strings. At the receiver, the channel output is decoded with respect to a bit-metric. An achievable rate is derived using random coding arguments. For the 8-ASK AWGN channel, numerical results show that at a spectral efficiency of 2 bits/s/Hz, the new scheme outperforms bit-interleaved coded modulation (BICM) without shaping and BICM with bit shaping (\text{i Fabregas} and Martinez, 2010) by 0.87 dB and 0.15 dB, respectively, and is within 0.0094 dB of the coded modulation capacity. The new scheme is implemented by combining a distribution matcher with a systematic binary low-density parity-check code. The measured finite-length gains are very close to the gains predicted by the asymptotic theory.
\end{abstract}

\section{Introduction}

\emph{Bit-interleaved coded modulation} (BICM) combines high order modulation with \emph{binary} error correcting codes\cite{zehavi1992psk}. This makes BICM attractive for practical application and BICM is widely used in standards, e.g., in DVB-T2/S2/C2. At a BICM receiver, \emph{bit-metric decoding} is used \cite[Sec.~II]{martinez2009bit}. We are \emph{not} considering BICM \emph{with iterative demapping-decoding} (BICM-ID, \cite{li1997bit}) here.

For bit-metric decoding, the signal points of a channel input constellation of size $2^m$ are labeled by bit strings of length $m$. The $m$ bit levels are treated independently at the decoder. We make this precise in Sec.~\ref{sec:rate}. Let $\vecB=(B_1,B_2,\dotsc,B_m)^T$ denote a column vector of $m$ binary random variables $B_i$, $i=1,2,\dotsc,m$, representing the bit levels. Conditioned on the channel input represented by $\vecB$, let $P_{Y|\vecB}$ be the distribution of the channel output $Y$. In \cite{caire1998bit}, it was shown that a bit-metric decoder achieves the rate
\begin{align}
\sum_{i=1}^m \miop(B_i;Y)\label{rate}
\end{align}
where $\miop(B_i;Y)$ denotes the mutual information of bit level $B_i$ and channel output $Y$. The proof in \cite{caire1998bit} assumes an ideal interleaver, which lets the $m$ bit levels see $m$ independent binary input channels. The authors in \cite{martinez2009bit} showed that \eqref{rate} is achievable without an ideal interleaver for independent and uniformly distributed $B_i$. We call this scheme \emph{uniform BICM}. In \cite{ifabregas2010bit}, the authors showed that \eqref{rate} is achievable for independent and arbitrarily distributed bit levels $B_i$. We call this scheme \emph{bit shaped BICM} (BS-BICM). It was illustrated in \cite{ifabregas2010bit} for the \emph{additive white Gaussian noise} (AWGN) channel that by properly choosing the bit level distributions, a shaping gain over uniform BICM can be achieved. The authors of \cite{bocherer2012efficient} proposed an algorithm to calculate the maximum rate of bit shaped BICM.
\begin{figure}
\footnotesize
\pgfplotsset{
width=\columnwidth,
height=0.3\textheight
}
\pgfplotsset{every axis title/.append style={at={(0.0,1.05)},
right
}}
%
%
\begin{tikzpicture}
\begin{axis}[
xmin=11.7,
xmax=12.8,
ymin=1.9,
ymax=2.1,
xlabel={\snr in dB},
ylabel={bits/s/Hz},
grid=both,
legend columns=1,
legend style={at={(0.01,0.99)},anchor=north west,draw=none},
legend entries = {$\frac{1}{2}\log_2(1+\snr)$, CM, SS-BMD, BS-BICM, uniform BICM},
y label style={at={(0.03,0.5)}},
]
\addplot[black,no markers]
table[x=snr,y=c]
{results/capacities.txt};
\addplot[magenta,no markers]
table[x=cm_E,y=cm]
{results/capacities.txt};
\addplot[blue,dashed,no markers]
table[x=ssbcm_E,y=ssbcm]
{results/capacities.txt};
\addplot[green,no markers]
table[x=bicm_E,y=bicm]
{results/capacities.txt};
\addplot[red,no markers]
table[x=bicm_uni_E,y=bicm_uni]
{results/capacities.txt};

\draw (axis cs:11.7,2) node {} -- (axis cs:12.8,2) node {};

\addplot[magenta,only marks]
table[x=cmsnr,y=cmrate]
{results/intuition.txt};
\addplot[blue,only marks]
table[x=ssbcmsnr,y=ssbcmrate]
{results/intuition.txt};
\addplot[green,only marks]
table[x=bicmsnr,y=bicmrate]
{results/intuition.txt};

\end{axis}
\end{tikzpicture}
\vspace{-0.3cm}
\caption{Power-rate curves for the $8$-ASK AWGN channel. The points on the curves that are marked by dots are discussed in Sec.~\ref{sec:intuition}.}
\vspace{-0.1cm}
\label{fig:capacities}
\end{figure}
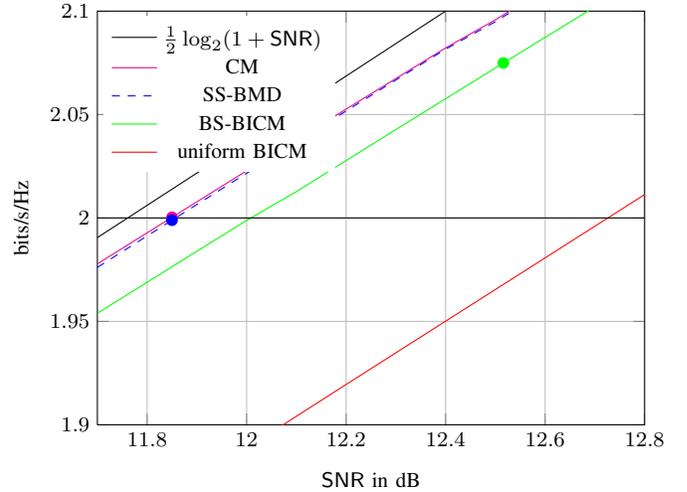

In Sec.~\ref{sec:rate}, we show that the rate
\begin{align}
\Bigl[\sum_{i=1}^m\miop(B_i;Y)\Bigr]-\left\{\Bigl[\sum_{i=1}^m\entop(B_i)\Bigr]-\entop(\vecB)\right\}\label{eq:ssbcmrate}
\end{align}
is achievable by a bit-metric decoder and for any \emph{joint} input distribution $P_{\vecB}$. $\entop$ denotes the entropy function. We call our scheme \emph{signal shaping with bit-metric decoding} (SS-BMD). In Sec.~\ref{sec:ask}, we calculate the power-rate curves for the AWGN channel with an equidistant 8-ASK input constellation. The resulting power-rate curves are displayed in Fig.~\ref{fig:capacities}. We find that at a spectral efficiency of 2 bits/s/Hz, SS-BMD outperforms uniform BICM and BS-BICM by 0.87 dB and 0.15 dB, respectively, and it lies within 0.0094 dB of the \emph{coded modulation} (CM) capacity. We then present in Sec.~\ref{sec:implementation} implementations of SS-BMD and BS-BICM where we combine a distribution matcher with the DVB-S2 rate 3/4 code \cite{etsi2009dvb}. To the best of our knowledge, no other implementation of BS-BICM has been reported in literature so far. \emph{We emphasize that we are considering \emph{probabilistic} shaping for BICM \emph{without iterative demapping-decoding}. Probabilistic shaping for BICM-ID has been considered in \cite{valenti2012constellation}, constellation shaping for BICM in \cite{hossain2010bicm} and constellation shaping for BICM-ID in \cite{legoff2007constellation}}. We compare to an implementation of uniform BICM with the DVB-S2 rate 2/3 code. At 2 bits/s/Hz and end-to-end block error probability of $10^{-2}$, SS-BMD outperforms uniform BICM and BS-BICM by 0.83 dB and 0.13 dB, respectively. SS-BMD operates within $0.9$ dB of the CM capacity and $0.99$ dB of the Shannon capacity.  In \cite{fischer1998combination}, similar results were achieved by combining probabilistic shaping with multilevel coding and multi stage decoding.

\section{Achievable Rate for SS-BMD}
\label{sec:rate}
Let $P_{Y|\vecB}$ be a \emph{discrete memoryless channel} (DMC) with input $\vecB=(B_1,B_2,\dotsc,B_m)^T$ and output $Y$. The $B_i$ are binary random variables.  
A \emph{bit-metric decoder} uses the metric
\begin{align}
q(\vecb,y)=\prod_{i=1}^m q_i(b_i,y)\label{eq:qmetric}
\end{align}
where $q_i(b_i,y)$ is a function of
\begin{align}
P_{B_iY}(b_i,y)=\sum_{\veca\in\{0,1\}^m\colon a_i=b_i}P_{Y|\vecB}(y|\veca)P_{\vecB}(\veca).\label{eq:pby}
\end{align}
If the $B_i$ are independent, then our definition is equivalent to \cite[Eq. (9)]{ifabregas2010bit}. If the $B_i$ are also uniformly distributed, then our definition is equivalent to \cite[Eq. (6)]{martinez2009bit}.

\begin{theorem}\label{theo:ssbcm}
For a DMC $P_{Y|\vecB}$ with finite input and output alphabets, the rate \eqref{eq:ssbcmrate} is achievable by a bit-metric decoder.
\end{theorem}
\begin{IEEEproof}
The proof is given in the appendix.
\end{IEEEproof}

\begin{figure}
\footnotesize
\centering
\begin{tikzpicture}
{
[start chain,node distance=0.4cm,every on chain/.style={join=by -},inode/.style={on chain,inner sep=0cm,minimum size=0.35cm}]
\node[inode]{};
\node[inode,draw,circle]{-7};
\node[inode,draw,circle]{-5} node[below=5pt]{$x_{\bsbm0\\0\\0\esbm}$};
\node[inode,draw,circle]{-3} node[below=5pt]{$x_{\bsbm0\\0\\1\esbm}$};
\node[inode,draw,circle]{-1} node[below=5pt]{$x_{\bsbm0\\1\\1\esbm}$};
\node[inode,draw,circle]{1} node[below=5pt]{$x_{\bsbm0\\1\\0\esbm}$};
\node[inode,draw,circle]{3} node[below=5pt]{$x_{\bsbm1\\1\\0\esbm}$};
\node[inode,draw,circle]{5} node[below=5pt]{$x_{\bsbm1\\1\\1\esbm}$};
\node[inode,draw,circle]{7} node[below=5pt]{$x_{\bsbm1\\0\\1\esbm}$};
\node[inode]{} node[below=5pt]{$x_{\bsbm1\\0\\0\esbm}$};
}
\end{tikzpicture}
\caption{The \emph{Binary Reflected Gray Code} as defined in \cite[Sec.~II.B]{agrell2011optimal}.}
\label{fig:gray}
\end{figure}
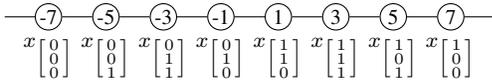
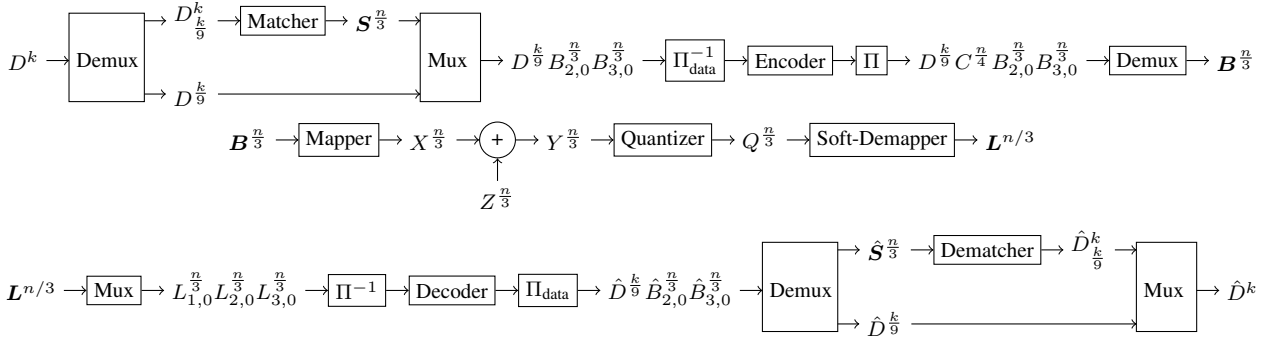
\begin{figure*}
\footnotesize
\centering
\begin{tikzpicture}
%
%
%
%
\node (matcher) [matrix of nodes,nodes in empty cells,column sep=0.3cm,inode/.style={anchor=center},nodes=inode] 
{
\node[text width=0.8cm] (mux1){}; &\node (a2){$D_\frac{k}{9}^k$};&\node(a3)[draw]{Matcher};&\node(a4){$\vecS^\frac{n}{3}$};&\node[text width=0.6cm](demux1){};\\
\node[text width=0.8cm] (mux2){Demux};&&&&\node[text width=0.6cm](demux2){Mux};\\
\node[text width=0.8cm] (mux3){};&\node (b2){$D^\frac{k}{9}$};&&&\node[text width=0.6cm](demux3){};\\
};

\node[draw,inner sep=0pt,text width=2cm](mux)[fit=(mux1) (mux2) (mux3)]{};
\node[draw,inner sep=0pt](demux)[fit=(demux1) (demux2) (demux3)]{};

{[start chain,every on chain/.style={join=by ->}]
\chainin (mux1);
\chainin (a2);
\chainin (a3);
\chainin (a4);
\chainin (demux1);
}
{[start chain,every on chain/.style={join=by ->}]
\chainin (mux3);
\chainin (b2);
\chainin (demux3);
}
%
%
%
%

%
%
%
%
{
[start chain=going left,node distance=0.3cm,every on chain/.style={join=by <-}]
\chainin (mux2);
\node[on chain]{$D^k$};
}
%
%
%
%
{
[start chain,node distance=0.3cm,every on chain/.style={join=by ->},inode/.style={on chain},nodes=inode]
\chainin (demux2);
\node{$D^\frac{k}{9}B_{2,0}^\frac{n}{3}B_{3,0}^\frac{n}{3}$};
\node[draw]{$\Pi^{-1}_\text{data}$};
\node[draw]{Encoder};
\node[draw]{$\Pi$};
\node{$D^\frac{k}{9}C^\frac{n}{4}B_{2,0}^\frac{n}{3}B_{3,0}^\frac{n}{3}$};
\node[draw]{Demux};
\node{$\vecB^\frac{n}{3}$};
}
\end{tikzpicture}
\begin{tikzpicture}
{
[start chain,node distance=0.3cm,every on chain/.style={join=by ->},inode/.style={on chain},nodes=inode]
\node{$\vecB^\frac{n}{3}$};
\node[draw]{Mapper};
\node{$X^\frac{n}{3}$};
\node[draw,circle]{+};
{[start branch=noise going below]}
\node{$Y^\frac{n}{3}$};
\node[draw]{Quantizer};
\node{$Q^\frac{n}{3}$};
\node[draw]{Soft-Demapper};
\node{$\vecL^{n/3}$};

{[continue branch=noise,every on chain/.style={join=by <-}]
\node[on chain]{$Z^\frac{n}{3}$};
}
}
\end{tikzpicture}
\begin{tikzpicture}
%
%
%
%
\node (matcher) [matrix of nodes,nodes in empty cells,column sep=0.3cm,inode/.style={anchor=center},nodes=inode] 
{
\node[text width=0.8cm] (mux1){}; &\node (a2){$\hat{\vecS}^\frac{n}{3}$};&\node(a3)[draw]{Dematcher};&\node(a4){$\hat{D}_\frac{k}{9}^k$};&\node[text width=0.6cm](demux1){};\\
\node[text width=0.8cm] (mux2){Demux};&&&&\node[text width=0.6cm](demux2){Mux};\\
\node[text width=0.8cm] (mux3){};&\node (b2){$\hat{D}^\frac{k}{9}$};&&&\node[text width=0.6cm](demux3){};\\
};

\node[draw,inner sep=0pt,text width=2cm](mux)[fit=(mux1) (mux2) (mux3)]{};
\node[draw,inner sep=0pt](demux)[fit=(demux1) (demux2) (demux3)]{};

{[start chain,every on chain/.style={join=by ->}]
\chainin (mux1);
\chainin (a2);
\chainin (a3);
\chainin (a4);
\chainin (demux1);
}
{[start chain,every on chain/.style={join=by ->}]
\chainin (mux3);
\chainin (b2);
\chainin (demux3);
}
%
%
%
%
{
[start chain=going left,node distance=0.3cm,every on chain/.style={join=by <-}]
\chainin (mux2);
\node[on chain]{$\hat{D}^\frac{k}{9}\hat{B}_{2,0}^\frac{n}{3}\hat{B}_{3,0}^\frac{n}{3}$};
\node[draw,on chain]{$\Pi_\text{data}$};
\node[draw,on chain]{Decoder};
\node[draw,on chain]{$\Pi^{-1}$};
\node[on chain]{$L_{1,0}^\frac{n}{3} L_{2,0}^\frac{n}{3} L_{3,0}^\frac{n}{3}$};
\node[draw,on chain]{Mux};
\node[on chain]{$\vecL^{n/3}$};
}
{
[start chain,node distance=0.3cm,every on chain/.style={join=by ->},inode/.style={on chain},nodes=inode]
\chainin (demux2);
\node[on chain]{$\hat{D}^k$};
}
\end{tikzpicture}
\caption{Transmitter and receiver of the SS-BMD system. \emph{Mux} is short for \emph{multiplexer} and \emph{Demux} is short for \emph{demultiplexer}.}
\label{fig:transceiver}
\end{figure*}
\section{$2^m$-ASK Modulation for the AWGN Channel}
\label{sec:ask}
The discrete time baseband AWGN channel is described by
\begin{align}
Y=X+Z\label{eq:ask:channel}
\end{align}
where $X$ and $Y$ are the input and output, respectively, and where $Z$ is zero mean and unit variance Gaussian noise. If the input is subject to an average power constraint $\snr$, the capacity of the AWGN channel is \cite[Theo.~7.4.2]{gallager1968information}
\begin{align}
\capacity(\snr)=\frac{1}{2}\log_2(1+\snr).
\end{align}
In practice, the input is restricted to a finite set $\setx$ of constellation points. We consider \emph{amplitude shift keying} (ASK) with $2^m$ equidistant constellation points, i.e., we have
\begin{align}
\setx_\text{ASK}=\{2\cdot i-2^m-1\colon i=1,2,\dotsc,2^m\}.
\end{align}
The points $x\in\setx_\text{ASK}$ are labeled by a binary vector $\vecB=(B_1,\dotsc,B_m)^T$. See Fig.~\ref{fig:gray} for an example with $m=3$. We model the channel input as $d\cdot x_{\vecB}$ where the label $\vecB$ is distributed according to the distribution $P_{\vecB}$ and where $d$ is a non-negative real number that scales the constellation. 
\subsection{Coded modulation}
The \emph{coded modulation} (CM) capacity of the ASK constellation is
\begin{align}
\begin{split}
\capacity_\mathrm{CM}(\snr)=&\max_{P_{\vecB},d}\miop(\bits;Y)\\
\st\quad&d^2\expop[x_{\vecB}^2]\leq\snr.
\end{split}
\end{align}
The maximization is both over the distribution $P_{\vecB}$ and the constellation scaling $d$.
\subsection{Bit Shaped BICM}
In BICM, the labeling of the constellation strongly influences the achievable rate, see \cite{agrell2011optimal}. We denote by $L$ the chosen labeling. An achievable rate for bit shaped BICM is by \cite[Eq.~(19)]{ifabregas2010bit}
\begin{align}
\begin{split}
\capacity_\mathrm{BICM}^L(\snr)=&\max_{P_{\bits},d}\sum_{i=1}^m\miop(B_i;Y)\\
\st\quad &P_{\bits} = \prod_{i=1}^{m}P_{B_i}\\
&d^2\expop[x_{\bits}^2]\leq\snr
\end{split}
\end{align}
The optimization is over the constellation scaling $d$ and the bit distributions $P_{B_i}$. When all bit distributions $P_{B_i}$ are uniform, we get the achievable rate of uniform BICM.
\subsection{SS-BMD}
\begin{theorem}
SS-BMD achieves the rate
\begin{align}
\begin{split}
&\hspace{-0.5cm}\capacity^L_\text{SS-BMD}(\snr)\\
&=\max_{P_{\bits},d}\sum_{i=1}^m\miop(B_i;Y)-\Bigl[\sum_{i=1}^m\entop(B_i)-\entop(\vecB)\Bigr]\\
&\text{\rm subject to}\quad d^2\expop[x_{\bits}^2]\leq\snr.\label{eq:ssbcmrate:awgn}
\end{split}
\end{align}
\end{theorem}
\begin{proof}
The theorem can be proven by adapting the proof of Theo.~\ref{theo:ssbcm} to the AWGN channel, similar to the approach taken in \cite[Sec.~III]{martinez2009bit}.
\end{proof}

\subsection{Numerical Results}
\label{subsec:8ask}

We evaluate achievable rates for $8$-ASK near a spectral efficiency of 2 bits/s/Hz. We choose the Gray labeling in Fig.~\ref{fig:gray}, since it is the best known labeling near 2 bits/s/Hz for BICM \cite[Fig.~2(b)]{agrell2011optimal}. We discretize the channel output $y$ into $2^9=512$ intervals; this choice effectively achieves the continuous CM capacity. We calculate the CM capacity by line search over $d$; for each value of $d$, maximization over $P_{\vecB}$ is a convex optimization problem. We use CVX \cite{cvx} to solve the problem. For bit shaped BICM, we again do a line search over $d$. For each value of $d$, maximization over $P_{B_1}P_{B_2}P_{B_3}$ is a non-convex optimization problem \cite{bocherer2012efficient}. We use the algorithm from \cite{bocherer2012efficient,website:bacm} to solve the problem. Uniform BICM is straight-forward. For SS-BMD, we use a heuristic. We evaluate the SS-BMD achievable rate in the values $d,P_{\vecB}$ that achieve the CM capacity. In Fig.~\ref{fig:capacities} we display the resulting power-rate curves.

\subsection{Discussion}
\label{sec:intuition}
The CM capacity of $8$-ASK is an upper bound for any $8$-ASK transmission scheme. At 2 bits/s/Hz, SS-BMD gets within 0.0094 dB of CM capacity while for BS-BICM, the gap to CM capacity is 0.16 dB. This is somewhat surprising, since we would expect that SS-BMD loses in terms of \emph{rate} over BS-BICM because of the correlated bit levels. In fact, this does happen, but SS-BMD gains in terms of \emph{SNR}. The dots in Fig.~\ref{fig:capacities} illustrate this. The blue and the red dot are obtained by evaluating the CM capacity and the SS-BMD rate in the same distribution $P_{\vecB}$. The green dot is obtained by evaluating the BS-BICM rate in the marginals $P_{B_1}$,$P_{B_2}$,$P_{B_3}$ of $P_{\vecB}$. Using the marginals leads to a rate gain of $0.076$ bits per channel use compared to SS-BMD, however, it also leads to an SNR loss of $0.67$ dB, which moves it away from the CM capacity curve. For all three dots, we used the same constellation scaling $d$.

\section{SS-BMD System Design}
\label{sec:system}

We design a system that lets us reap the shaping gap between SS-BMD and uniform BICM. The distribution $P_{\vecB}$ that achieves CM capacity is symmetric around zero. The first bit level of Gray labeling chooses the sign of the constellation point, see Fig.~\ref{fig:gray}. The labeling of the other bit levels is symmetric around zero. This means that $B_1$ is stochastically independent of $B_2\dotsb B_m$, i.e., we have
\begin{align}
P_{\vecB}(\vecb)=P_{B_1}(b_1)P_{B_2\dotsb B_m}(b_2\dotsb b_m),\; \forall\vecb\in\{0,1\}^m.
\end{align}
Furthermore, $B_1$ is uniformly distributed. Systematic binary encoders copy data bits to the codeword and append parity bits. Thus, if the data bits have a non-uniform distribution, \emph{this distribution is preserved by the encoder}. However, the parity bits are modulo 2 sums of data bits, so their distribution is approximately uniform and it is reasonable to model the parity bits to be independent and uniformly distributed \cite[Chap.~7]{bocherer2012capacity},\cite{bocherer2011operating}. An SS-BMD system could mimic the capacity-achieving distribution as follows.  Use a distribution matcher on data bits to generate $B_2\dotsb B_m$ according to $P_{B_2\dotsb B_m}$. Encode $B_2\dotsb B_m$ by a systematic encoder and use the parity bits and possibly additional data bits for bit level 1. This is possible as long as the coding rate of the code fulfills
\begin{align}
\frac{k}{n}\geq\frac{m-1}{m}.\label{eq:coderatecond}
\end{align}
Ungerb\"ock  \cite{ungerbock1982channel} made the observation that for reliable transmission over the AWGN channel, it suffices to add one bit of
redundancy per real dimension. This observation was analytically confirmed in \cite{ozarow1990capacity} and it was experimentally confirmed in \cite{fischer1998combination}, for example. This suggests that condition \eqref{eq:coderatecond} is feasible.

\section{Implementation}
\label{sec:implementation}

We now implement the SS-BMD system that we outlined in Sec.~\ref{sec:system}. Our target is to transmit $2$ bits per channel use reliably over the AWGN channel with the Gray labeled 8-ASK input constellation, as described in Sec.~\ref{subsec:8ask}. We use the DVB-S2 LDPC codes \cite{etsi2009dvb} with block length $n=64800$ and code rate $3/4$ for SS-BMD and BS-BICM, and with code rate $2/3$ for uniform BICM. We next discuss the important parts of our SS-BMD system. A complete flow chart is provided in Fig.~\ref{fig:transceiver}.

\subsection{Notation}

For row vectors, we use the notation
\begin{align}
V_{i}^j=(V_{i+1},V_{i+2},\dotsc,V_{j}).
\end{align}
If it is clear from the context, we write $V^j$ instead of $V_0^j$. The vector $V_i^j$ thus has $j-i$ entries and we can write $V^j=(V_0^i,V_i^j)$ for $0< i< j$.

We denote column vectors using a bold font. The labels of $n$ consecutive ASK-8 signal points are denoted by
\begin{align*}
\vecB^n=(\vecB_1,\vecB_2,\dotsc,\vecB_n),\quad \vecB_i=\bbm B_{1,i}\\B_{2,i}\\B_{3,i}\ebm, i=1,2,\dotsc,n.
\end{align*}
Bit levels 2 and 3 together are denoted by $\vecS$, i.e., we have
\begin{align}
\vecS=\bbm B_2\\B_3\ebm,\quad\vecB=\bbm B_1\\\vecS\ebm,\quad\vecB^n=\bbm B_{1,0}^n\\\vecS^n\ebm.
\end{align}

\subsection{Adapt the Signal Point Distribution to the Code Rate}

For the Gray labeling in Fig.~\ref{fig:gray}, the distribution that achieves the SS-BMD rate of $2$ bits/s/Hz is $P^*_{\vecB}=P^*_{B_1}P^*_{\vecS}$ with $P^*_{B_1}(0)=P^*_{B_1}(1)=1/2$ and
\begin{align}
&P^*_{\vecS}(00)=0.0579\\
&P^*_{\vecS}(01)=0.1507\\
&P^*_{\vecS}(11)=0.3237\\
&P^*_{\vecS}(10)=0.4676.
\end{align}
We use a rate 3/4 code, so according to our outline in Sec.~\ref{sec:system}, we use ``matched data'' with joint distribution $P_{\vecS}$ on bit levels 2 and 3. We use $1/4$ of bit level $1$ for uniformly distributed data, and we use the remaining $3/4$ of bit level $1$ for parity bits. This bit level assignment is visualized in Fig.~\ref{fig:bitlevels}. The distribution $P^*_{B_1}P^*_{\vecS}$ results in
\begin{align}
\frac{1}{4}\entop(P^*_{B_1})+\entop(P^*_{\vecS})=0.25+1.6891=1.9391.
\end{align}
We therefore need to choose a distribution for $\vecS$ that is close to $P^*_{\vecS}$, but whose entropy is equal to $1.75$. We choose
\begin{align}
P_{\vecS}=&\argmin_{P}\kl(P\Vert P^*_{\vecS})\nonumber\\
&\st\;\entop(P)\geq 1.75\label{eq:id}
\end{align}
where $\kl(P\Vert P^*_{\vecS})$ denotes the \emph{informational divergence} or \emph{relative entropy} of $P$ and $P^*_{\vecS}$ \cite[Sec.~2.3]{cover2006elements}. The optimization problem \eqref{eq:id} is convex and by solving the KKT conditions \cite[Sec. 5.5.3]{boyd2004convex}, we find the solution
\begin{align}
P_{\vecS}(\veca)=\frac{P^*_{\vecS}(\veca)^\lambda}{\displaystyle\sum_{\vecb\in\{0,1\}^2}P^*_{\vecS}(\vecb)^\lambda}
\end{align}
where $\lambda$ is chosen such that $\entop(P_{\vecS})=1.75$. We find $\lambda=0.8672$ and
\begin{align}
&P_{\vecS}(00)=0.0722\\
&P_{\vecS}(01)=0.1654\\
&P_{\vecS}(11)=0.3209\\
&P_{\vecS}(10)=0.4415.
\end{align}
Since we have fixed the distribution to $P_{\vecS}$, the SNR and the achievable rate depend only on the constellation scaling $d$. We choose $d$ such the SS-BMD rate evaluates to $2$ bits/s/Hz. We observe a loss of 0.028 dB as compared to $P^*_{\vecS}$. 

\subsection{Matcher Input and Output Lengths}
\begin{figure}
\footnotesize
\centering
\begin{tikzpicture}
\draw (0,0) node{bit level 3};
\draw (0.65,0) node[minimum width = 7.5cm,draw,anchor=west,minimum height=0.6cm]{matched data bits $B_{3,0}^\frac{n}{3}$};
\draw (0,1) node{bit level 2};
\draw (0.65,1) node[minimum width = 7.5cm,draw,anchor=west,minimum height=0.6cm]{matched data bits $B_{2,0}^\frac{n}{3}$};
\draw (0,2) node{bit level 1};
\draw (0.65,2) node[minimum width = 1.875cm,draw,anchor=west,minimum height=0.6cm]{data bits $D^\frac{n}{12}$};
\draw (2.525,2) node[minimum width = 5.625cm,draw,anchor=west,minimum height=0.6cm]{parity bits $C^\frac{n}{4}$};
\end{tikzpicture}
\caption{Visualization of the bit levels.}
\label{fig:bitlevels}
\end{figure}
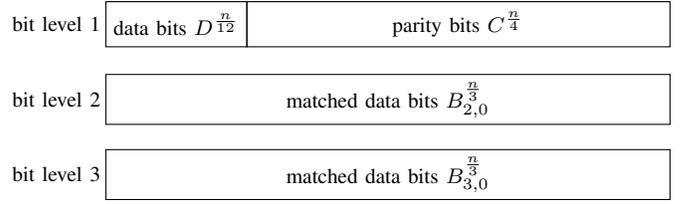
At the system input, we have $k=\frac{2n}{3}=43200$ data bits $D^k$ that are independent and uniformly distributed. For bit level 1, we use $\frac{n}{12}=\frac{k}{9}=5400$ data bits $D^\frac{k}{9}$ according to Fig.~\ref{fig:bitlevels}. For bit levels 2 and 3, we map the remaining $8k/9=37800$ data bits $D_\frac{k}{9}^k$ to $\frac{n}{3}=21600$ symbols that are distributed according to $P_{\vecS}$.

We verify that this approach is in accordance with the distribution $P_{\vecS}$. An ideal distribution matcher performs a one-to-one mapping from the input to the output. Thus, information is conserved and we have
\begin{align}
\frac{n}{3}\cdot\entop(P_{\vecS})=21600\cdot 1.75 = 37800=\frac{8k}{9}.
\end{align}

\subsection{Interleaver}
\label{sec:interleaver}
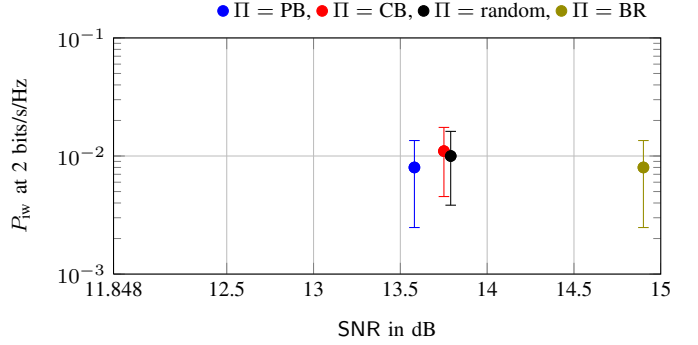
\begin{figure}
\footnotesize
\centering
\pgfplotsset{
width=\columnwidth,
height=0.2\textheight
}
\pgfplotsset{every axis title/.append style={at={(0.0,1.05)},
right
}}
%
%
\begin{tikzpicture}
\begin{semilogyaxis}[
xmin=11.848,
xmax=15,
ymin=0.001,
ymax=0.1,
xtick = {11.848,12.5,13,13.5,14,14.5,15},
xticklabels = {11.848,12.5,13,13.5,14,14.5,15},
xlabel={\snr in dB},
ylabel={$\perr$ at 2 bits/s/Hz},
grid=major,
legend columns=4,
legend style={at={(1.0,1.03)},anchor=south east,draw=none},
legend entries = {$\Pi=\text{PB,}$,$\Pi=\text{CB,}$,$\Pi=\text{random,}$,$\Pi=\text{BR}$},
]
\addplot[blue,only marks,
mark=*,
error bars/.cd,y dir=plus,y explicit]
table[x=snr,y=piw,y error=U]
{results/321.txt};
\addplot[red,only marks,
mark=*,
error bars/.cd,y dir=plus,y explicit]
table[x=snr,y=piw,y error=U]
{results/CB.txt};
\addplot[black,only marks,
mark=*,
error bars/.cd,y dir=plus,y explicit]
table[x=snr,y=piw,y error=U]
{results/RI.txt};
\addplot[olive,only marks,
mark=*,
error bars/.cd,y dir=plus,y explicit]
table[x=snr,y=piw,y error=U]
{results/123.txt};

\addplot[blue,only marks,no markers,
error bars/.cd,y dir=minus,y explicit]
table[x=snr,y=piw,y error=L]
{results/321.txt};
\addplot[red,only marks,no markers,
error bars/.cd,y dir=minus,y explicit]
table[x=snr,y=piw,y error=L]
{results/CB.txt};
\addplot[black,only marks,no markers,
error bars/.cd,y dir=minus,y explicit]
table[x=snr,y=piw,y error=L]
{results/RI.txt};
\addplot[olive,only marks,no markers,
error bars/.cd,y dir=minus,y explicit]
table[x=snr,y=piw,y error=L]
{results/123.txt};


\end{semilogyaxis}
\end{tikzpicture}
\vspace{-0.3cm}
\caption{SNR versus $\perr=\Pr\{D^k\neq\hat{D}^k\}$ at 2 bits/s/Hz for uniform BICM. The DVB-S2 rate 2/3 LDPC code is used with the different interleavers discussed in Sec.~\ref{sec:interleaver}. 11.848 dB corresponds to the CM capacity.}
\label{fig:pi}
\end{figure}
Since the DVB-S2 codes are systematic and highly structured \cite{etsi2009dvb}, the performance may depend on which coded bits are used for which bit level. This is controlled by the interleaver $\Pi$, see Fig.~\ref{fig:transceiver}.
Since we want to compare to uniform BICM, we choose an interleaver that works well for uniform BICM with the DVB-S2 rate 2/3 code. We adjust the SNR such that the information word error probability
\begin{align}
\perr=\Pr\{D^k\neq\hat{D}^k\}
\end{align}
is around $\perr=10^{-2}$. We try a random interleaver, the \emph{consecutive-bit} (CB) interleaver \cite{li2005bit}, the \emph{bit-reliability} (BR) interleaver \cite{li2005bit}, and an interleaver that is defined by the following permutation of the codeword $V^n$:
\begin{align}
B_{1,0}^\frac{n}{3}=&V_{43200}^{64800}\label{eq:cb1}\\
B_{2,0}^\frac{n}{3}=&V_{21600}^{43200}\\
B_{3,0}^\frac{n}{3}=&V_0^{21600}\label{eq:cb3}.
\end{align}
For the DVB-S2 codes, the parity bits are appended to the data bits. The interleaver defined in \eqref{eq:cb1}--\eqref{eq:cb3} copies the parity bits to bit level 1. We therefore call it the \emph{parity-bit} (PB) interleaver. The simulation results for $\perr\approx 10^{-2}$ are shown in Fig.~\ref{fig:pi}. The PB interleaver performs best, and we therefore use it in our implementation. Note that we made this choice based on simulation results for a specific LDPC code, namely the rate 2/3 DVB-S2 code. We do not know if the PB interleaver is the optimal choice and other codes may lead to other choices.

We use the PB interleaver also for BS-BICM and SS-BMD. To preserve the bit level assignment of the data bits as shown in Fig.~\ref{fig:bitlevels}, we apply the inverse of $\Pi$ to the data bits before encoding them, specifically, we apply $\Pi^{-1}_\text{data}$, which is given by
\begin{align}
V_0^{21600}&=B_{3,0}^\frac{n}{3}\\
V_{21600}^{43200}&=B_{2,0}^\frac{n}{3}\\
V_{43200}^{48600}&=D^\frac{k}{9}.
\end{align}
Since the PR interleaver copies the parity bits $C^\frac{n}{4}$ to bit level 1, the chain $\Pi^{-1}_\text{data}$, rate 3/4 DVB-S2 encoder and $\Pi$ realizes the bit level assignment in Fig.~\ref{fig:bitlevels}.

\subsection{Decoder}
\begin{figure}
\footnotesize
\centering
\pgfplotsset{
width=\columnwidth,
height=0.25\textheight
}
\pgfplotsset{every axis title/.append style={at={(0.0,1.05)},
right
}}
%
%
\begin{tikzpicture}
\begin{semilogyaxis}[
xmin=11.848,
xmax=14,
ymin=0.0003,
ymax=0.3,
xlabel={\snr in dB},
ylabel={$\perr$ at 2 bits/s/Hz},
grid=both,
xtick = {11.848,12.5,13,13.5,14},
xticklabels={11.848,12.5,13,13.5,14},
extra x ticks={12.1,12.2,12.3,12.4,12.6,12.7,12.8,12.9,13.1,13.2,13.3,13.4,13.6,13.7,13.8,13.9},
extra x tick labels={},
legend columns=4,
legend style={at={(1.0,1.03)},anchor=south east,draw=none},
legend entries = {\text{SS-BMD},\text{BS-BICM},\text{uniform BICM}},
]
\addplot[blue,only marks,
mark=*,
error bars/.cd,y dir=plus,y explicit]
table[x=snr,y=pe,y error=dev]
{results/sserrors.txt};
\addplot[green,only marks,
mark=*,
error bars/.cd,y dir=plus,y explicit]
table[x=snr,y=piw,y error=U]
{results/errorshaped.txt};
\addplot[red,only marks,
mark=*,
error bars/.cd,y dir=plus,y explicit]
table[x=snr,y=piw,y error=U]
{results/erroruniform.txt};

\addplot[blue,only marks,no markers,
error bars/.cd,y dir=minus,y explicit]
table[x=snr,y=pe,y error=dev]
{results/sserrors.txt};
\addplot[green,only marks,no markers,
error bars/.cd,y dir=minus,y explicit]
table[x=snr,y=piw,y error=L]
{results/errorshaped.txt};
\addplot[red,only marks,no markers,
error bars/.cd,y dir=minus,y explicit]
table[x=snr,y=piw,y error=L]
{results/erroruniform.txt};


\end{semilogyaxis}
\end{tikzpicture}
\vspace{-0.3cm}
\caption{SNR versus $\perr=\Pr\{D^k\neq\hat{D}^k\}$ at 2 bits/s/Hz. The PB interleaver (see Sec.~\ref{sec:interleaver}) is used. For uniform BICM, the DVB-S2 rate 2/3 LDPC code is employed. For SS-BMD and BS-BICM, an arithmetic distribution matcher \cite{website:adm} and the DVB-S2 rate 3/4 LDPC code is used.  11.848 dB corresponds to the CM capacity.}
\label{fig:piw}
\end{figure}
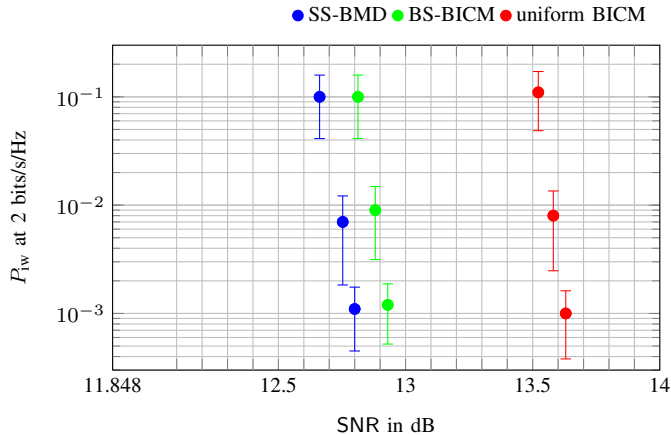
We consider one channel use. The bit label $\vecB$ gets mapped to the signal point $X=x_{\vecB}$ at the input of the AWGN channel \eqref{eq:ask:channel}. The output $Y=X+Z$ is quantized by a $5$ bit quantizer, and the soft-demapper uses the quantizer output $Q$ to calculate the soft information for each bit level $i=1,2,3$. The soft information is 
\begin{align}
I_i=L_i+\pi_i
\end{align}
where $L_i$ and $\pi_i$ are the log-likelihood ratio and the a-priori information, respectively. The log-likelihood ratios are calculated as
\begin{align}
L_i = \ln\frac{P_{Q|B_i}(Q|0)}{P_{Q|B_i}(Q|1)}
\end{align}
where
\begin{align}
P_{Q|B_i}(Q|a)=&\sum_{\vecb\in\{0,1\}^3\colon b_i=a}\frac{P_{Q B_1\vecS}(Q\vecb)}{P_{B_i}(a)}
\end{align}
with the marginals $P_{B_1}$ and
\begin{align}
&P_{B_2}(a)=\sum_{b\in\{0,1\}}P_{\vecS}(ab)\\
&P_{B_3}(a)=\sum_{b\in\{0,1\}}P_{\vecS}(ba).
\end{align}
The a-priori information for bit level $i$ is
\begin{align}
\pi_i=\ln\frac{P_{B_i}(0)}{P_{B_i}(1)}.
\end{align}

\subsection{Numerical Results}

For SS-BMD and BS-BICM, we use an arithmetic distribution matcher with controllable overflow \cite{website:adm}. For a detailed discussion of the arithmetic matcher, we refer the reader to the extended version of this paper, which is in preparation. The results for SS-BMD, BS-BICM and uniform BICM are displayed in Fig.~\ref{fig:piw}. We observe that the SNR gains of SS-BMD and BS-BICM over uniform BICM fit well to the asymptotic gains promised by the power-rate curves in Fig.~\ref{fig:capacities}. This suggests that we could effectively remove the entire shaping gap by using SS-BMD.


\section{Conclusions}

In this work, we showed that the CM capacity of the AWGN channel can effectively be achieved by a bit-metric decoder without iterative demapping-decoding. Our simulation results show that by combining Gray labeling, a distribution matcher and a binary code with a systematic encoder, the shaping gap can be removed. The remaining gap to capacity is because of the imperfections of the employed code. We observed that the interleaver has a strong impact on the performance. This suggests that interleaver and code should be designed together taking the varying reliabilities of the bit levels into account. To achieve very low error probabilities, LDPC codes with low error floor should be combined with zero error distribution matchers. The design of such distribution matchers is part of our current research. 	The techniques developed in this work may be useful in other scenarios, such as peak power constraints, fading channels, multiple antennas, and multiple users. 

\section{Acknowledgment}

The author would like to thank L. Barletta, M. Stinner and G. Kramer for support in this work. This work was supported by the German Ministry of Education and Research in the framework of an Alexander von Humboldt Professorship.

\appendix

\section{Proof of Theorem \ref{theo:ssbcm}}

We prove the theorem by random coding arguments. We use letter typicality as defined in \cite[Chap.~1]{kramer2008topics}. $\sett_\epsilon^n(P_{\vecB Y})$ is the set of sequences $\vecb^n$, $y^n$ that are jointly $\epsilon$-typical with respect to $P_{\vecB Y}$. The set of conditionally typical sequences is defined as
\begin{align}
\sett_\epsilon^n(P_{\vecB Y}|y^n):=\{\vecb^n\colon (\vecb^n,y^n)\in\sett^n_\epsilon(P_{\vecB Y})\}.
\end{align}

\emph{Codebook Construction:}
Choose $2^{nR}$ codewords of length $n$ by choosing the $n\cdot 2^{nR}$ symbols according to $P_{\vecB}$. Denote the resulting codebook by $\setc$.

\emph{Encoding:} Given message $w\in[1:2^{nR}]$, transmit $\vecb^n(w)$.

\emph{Decoding:} For $\epsilon_1>\epsilon\geq 0$, we define the bit metric
\begin{align}
q_i(b_i^n,y^n)=\begin{cases}
1,&b_i^n\in\sett^n_{\epsilon_1}(P_{B_iY}|y^n)\\
0,&\text{otherwise}.
\end{cases}
\end{align}
The corresponding decoding metric is
\begin{align}
q(\vecb^n,y^n)=\prod_{i=1}^m q_i(b_i^n,y^n).
\end{align}
We define the set
$
\hat{\setb}(y^n):=\{\vecb^n\in\setc\colon q(\vecb^n,y^n)=1\}.
$
The decoder output is
\begin{align}
\begin{cases}
\vecb^n,&\text{if }\setb(y^n)=\{\vecb^n\}\\
\text{error},&\text{otherwise}.
\end{cases}
\end{align}

\emph{Analysis:} Suppose message $w$ was encoded. The two error events are
\begin{align}
&\mathcal{E}_1:=\{\vecB^n(w)\notin\hat{\setb}(Y^n)\}\\
&\mathcal{E}_2:=\{\exists\tilde{w}\neq w\colon\vecB^n(\tilde{w})\in\hat{\setb}(Y^n)\}.
\end{align}
\emph{First error event:} The random experiment for $\mathcal{E}_1$ has the distribution $P_{\vecB Y}^n$. We have
\begin{align}
\Pr&(\mathcal{E}_1)=1-\Pr[q(\vecB^n,Y^n)=1]\nonumber\\
&= 1-\Pr\left[\bigcap_{i=1}^m \{B_i^n\in\sett_{\epsilon_1}^n(P_{B_iY}|Y^n)\}\right]\nonumber\\
&\oleq{a}1-\Pr\left[(\vecB^n,Y^n)\in\sett^n_{\epsilon_1}(P_{\vecB Y})\right]
\overset{\text{(b)}}{\overset{n\to\infty}{\to}}0\label{eq:rate:e1}
\end{align}
where (a) follows because joint typicality implies marginal typicality \cite[Sec.~1.5]{kramer2008topics}. The limit (b) follows by \cite[Theo.~1.1]{kramer2008topics}.

\emph{Second error event:}
The random experiment for $\mathcal{E}_2$ has the distribution $P_{\vecB}^nP_Y^n$. The probabilities of $Y^n\in\sett^n_\epsilon(P_Y)$ and $\vecB^n\in\sett_\epsilon^n(P_{\vecB})$ approach $1$ as $n\to\infty$, by \cite[Theo.~1.1]{kramer2008topics}. It therefore suffices to analyze for $y^n\in\sett_\epsilon^n(P_Y)$ the probability
\begin{align}
\Pr[\mathcal{E}_2|Y^n=y^n,\vecB^n\in\sett_\epsilon^n(P_{\vecB})].
\end{align}
By \cite[Theo.~1.2]{kramer2008topics}, we have
\begin{align}
|\sett^n_{\epsilon_1}(P_{B_iY}|y^n)|\leq 2^{n\entop(B_i|Y)(1+\epsilon_1)}.
\end{align}
The size of $\hat{B}(y^n)$ is thus bounded as
\begin{align}
|\hat{B}(y^n)|\leq 2^{n\sum_{i=1}^m \entop(B_i|Y)(1+\epsilon_1)}.\label{eq:rate:sizeB}
\end{align}
By \cite[Eq. (1.10) \& (1.12)]{kramer2008topics}, we have
\begin{align}
\Pr[\vecB^n=\vecb^n|\vecB^n\in\sett_\epsilon^n(P_{\vecB})]\leq \frac{2^{-n\entop(\vecB)(1-\epsilon)}}{1-\delta_\epsilon(n)}\label{eq:bound}
\end{align}
where $\delta_\epsilon(n)\overset{n\to\infty}{\to}0$. We assume $n$ is large enough such that $\delta_\epsilon(n)\leq 1/2$. The bound \eqref{eq:bound} then becomes
\begin{align}
\Pr[\vecB^n=\vecb^n|\vecB^n\in\sett_\epsilon^n(P_{\vecB})]\leq 2\cdot2^{-n\entop(\vecB)(1-\epsilon)}\label{eq:bound1}
\end{align}
We have
\begin{align}
&\Pr[\mathcal{E}_2|Y^n=y^n,\vecB^n\in\sett_\epsilon^n(P_{\vecB})]\nonumber\\
&\leq (2^{nR}-1)\sum_{\vecb^n\in\hat{\setb}(y^n)}\Pr[\vecB^n=\vecb^n|\vecB^n\in\sett_\epsilon^n(P_{\vecB})]\nonumber\\
&\oleq{a} 2^{nR}\sum_{\vecb^n\in\hat{\setb}(y^n)}2\cdot2^{-n\entop(\vecB)(1-\epsilon)}\nonumber\\
&\oleq{b} 2^{nR}2^{n\sum_{i=1}^m\entop(B_i|Y)(1+\epsilon_1)}\cdot2\cdot2^{-n\entop(\vecB)(1-\epsilon)}\label{eq:e2bound}
\end{align}
where (a) follows by \eqref{eq:bound1} and where we used \eqref{eq:rate:sizeB} in (b). The term in \eqref{eq:e2bound} goes to zero for $n\to\infty$ if
\begin{align}
&R+\Bigl[\sum_{i=1}^m\entop(B_i|Y)(1+\epsilon_1)\Bigr]-\entop(\vecB)(1-\epsilon)<0.\label{eq:rate:condition}
\end{align}
Using $\sum_{i=1}^m\entop(B_i|Y)\leq m$ and $\entop(\vecB)\leq m$, we have
\begin{align}
&R<\sum_{i=1}^m\bigl[\entop(B_i)-\entop(B_i|Y)\bigr]\nonumber\\
&\hspace{2cm}-\Bigl\{\Bigl[\sum_{i=1}^m \entop(B_i)\Bigr]-\entop(\vecB)\Bigr\}-m(\epsilon_1+\epsilon)\nonumber\\
&\Leftrightarrow R<\sum_{i=1}^m\miop(B_i;Y)\nonumber\\
&\hspace{1.5cm}-\Bigl\{\Bigl[\sum_{i=1}^m \entop(B_i)\Bigr]-\entop(\vecB)\Bigr\}-m(\epsilon_1+\epsilon)\label{eq:rate:e2}
\end{align}
for any $0<\epsilon<\epsilon_1$. Thus, for any rate $R$ less than \eqref{eq:ssbcmrate}, it follows by \eqref{eq:rate:e1} and \eqref{eq:rate:e2} that the probability of decoding error can be made as small as desired by choosing $n$ large enough.

\bibliographystyle{IEEEtran}
\normalsize
\bibliography{IEEEabrv,confs-jrnls,references}

\end{document}